\documentclass{epl}

\title{Stability of undissociated screw dislocations in zinc-blende covalent materials from first principle simulations}
\shorttitle{Screw calculations in zinc-blende materials}
\author{L. Pizzagalli\thanks{E-mail: \email{Laurent.Pizzagalli@univ-poitiers.fr}} \and P. Beauchamp \and J. Rabier}
\institute{
  Laboratoire de M\'etallurgie Physique, CNRS UMR 6630, Universit\'e
de Poitiers,  B.P~30179, F-86962~Futuroscope Chasseneuil Cedex, France
}
\pacs{61.72.Lk}{Linear defects: dislocations, disclinations}
\pacs{31.15.Ar}{Ab initio calculations}
\pacs{61.82.Fk}{Semiconductors}

\begin{document}

\maketitle

\begin{abstract}
The properties of perfect screw dislocations have been investigated for several zinc-blende materials such as diamond, Si, $\beta$-SiC, Ge and GaAs, by performing first principles calculations. For almost all elements, a core configuration belonging to shuffle set planes is favored, in agreement with low temperature experiments. Only for diamond, a glide configuration has the lowest defect energy, thanks to an sp$^2$ hybridization in the core. 

\end{abstract}

\begin{figure}
\onefigure[width=10cm]{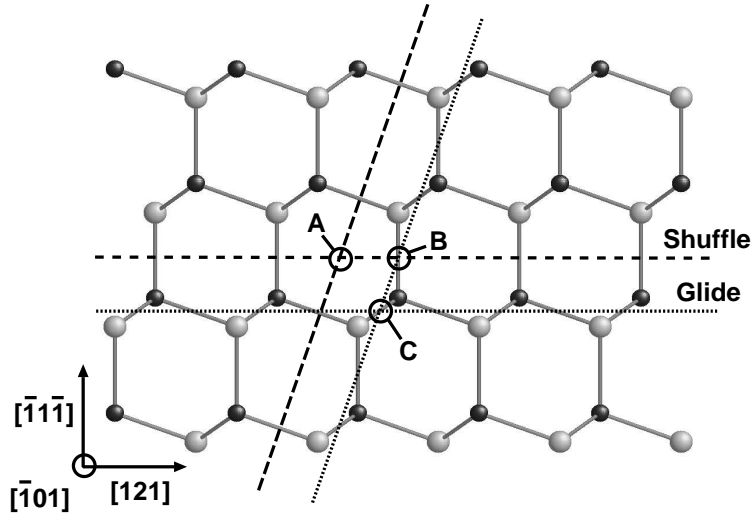}
\caption{Ball-and-stick representation of the zinc-blende structure ($(\bar{1}01)$ plane). The A, B and C circles indicate three possible positions of the dislocation line. Broken (dotted) lines show the 'shuffle' ('glide') \{111\} planes.}
\label{shuffleglide}
\end{figure}

Several technologically interesting covalent materials are semiconductors or insulators with a zinc-blende crystalline structure. The most famous is silicon, extensively used in electronic devices. It is usually considered as a model for other materials with the same structure, and most of its properties are well known, thanks to an impressive number of dedicated studies. For instance, the plasticity of silicon has been largely investigated both experimentally and theoretically. At high temperature, silicon is ductile, and  dislocations are dissociated into Shockley partials; since the dislocations are frequently aligned along the $\langle110\rangle$ dense directions, they mostly appear as 30$^\circ$ and 90$^\circ$ partials and slip in the so-called 'glide' set of \{111\} planes  (Fig.~\ref{shuffleglide}) \cite{Hir82WIL}. Calculations have allowed to determine the core structure of partials \cite{Due91PRB,Big92PRL,Leh99EMIS}, although the most stable configuration of the 90$^\circ$ is still not known with certainty \cite{Ben97PRL,Leh98PRL,Nun00JPCM,Mir03PRB}. At low temperature, in the brittle regime, surface scratch tests and confining pressure experiments have  recently revealed the presence of undissociated dislocations, with screw, 60$^\circ$, 30$^\circ$ and 41$^\circ$ orientations\cite{Rab00JPCM,Rab01MSE}, so that very little investigation of these perfect dislocations recently have been reported. So far, theoretical studies have shown that the non-dissociated screw belongs to the 'shuffle' set, though there has been a controversy about the most stable structure \cite{Koi00PMA,Piz03PMA}. 

Other covalent materials with zinc-blende structure are expected to behave like silicon, at least at high temperature. Partial dislocations have been theoretically characterized for $\beta$-SiC (cubic 3C-SiC) \cite{Sit95PRB,Blu02JPCM,Blu03PRB}, diamond \cite{Bla00PRL,Blu03JPCM,Blu02PRB}, Ge \cite{Nun00JPCM,Mir03PRB,Nun98PRB}, 
GaAs \cite{Obe95PRB,Bec02JPCM}, and others \cite{Jus01SSC,Jus00JPCM}. However, it is not clear whether the results obtained at low temperature for silicon remained valid for those materials, and little is known on that subject. Experiments have shown that low temperature deformation of GaAs yields undissociated screw dislocations \cite{Suz99PMA}, like in silicon, and recent calculations by Blumenau et al focussed on perfect screw in diamond \cite{Blu03PRB}. Recently, we have investigated the properties of undissociated screw dislocations in several zinc-blende materials \cite{Piz02JPCM}. 
Our simulations, performed with empirical potentials, indicated stable configurations in the shuffle set for Si and Ge, and in the glide set for $\beta$-SiC and diamond. One has to be cautious with these results though, since empirical potentials may fail to describe correctly the energetics of the highly distorted core geometry \cite{Piz03PMA}. This is especially true in the case of SiC or GaAs, for which charge transfers occur. 

In this paper, we report first principles calculations of undissociated screw dislocations in several zinc-blende materials, namely Ge, Si, $\beta$-SiC, diamond and GaAs. The theoretical approach is very similar to the one proposed in~\cite{Piz03PMA}: a quadrupolar arrangement of screw dislocations is built using 
anisotropic elasticity theory, from which a $12\times12\times1$ cell including two dislocations with opposite Burgers vectors is cut. Periodic boundary conditions are applied to recover the quadrupolar arrangement \cite{Big92PRL}. After relaxation, quantities attached to a single dislocation are obtained by removing contributions from dislocation  interactions, determined from anisotropic elasticity theory \cite{Piz03PMA}. Density functional theory calculations have been performed with the ABINIT code \cite{ABINIT}. Norm-conserving pseudopotentials \cite{Tro91PRB}, local density approximation, and two special k-points along the dislocation line have been used. We employed plane-waves energy cut-off of 10~Ry for Si, Ge and GaAs, and 40~Ry for SiC and diamond. These parameters allow a satisfactory determination of 
elastic constants in all materials (Tab.~\ref{tabconst}). 

\begin{table}
\caption{Elastic constants, in Mbar, calculated for several materials and compared to experiments (in parentheses).}\label{tabconst}
\begin{center}

\begin{tabular}{@{\extracolsep{10pt}}*{6}{c}}
\hline
\hline
  & C & SiC  & Si & Ge & GaAs \\
\hline
B         & 4.52 (4.42)   & 2.21 (2.25)  & 0.99 (0.99)  & 0.77 (0.77) & 0.80 (0.75)   \\
C$_{11}$  & 10.81 (10.79) & 3.96 (3.90)  & 1.64 (1.67)  & 1.32 (1.29) & 1.27 (1.19)   \\
C$_{21}$  & 1.37 (1.24)   & 1.34 (1.42)  & 0.66 (0.65)  & 0.50 (0.48) & 0.56 (0.53)   \\
C$_{44}$  & 5.92 (5.78)   & 2.54 (2.56)  & 0.78 (0.81)  & 0.67 (0.67) & 0.64 (0.60)   \\
\hline
\hline
\end{tabular}
\end{center}\end{table}

Three possible dislocation line positions have been considered in this work (Fig.~\ref{shuffleglide}). A is in the center of the projected hexagon and is at the intersection of two shuffle planes. Position B is at the center of a long hexagon bond and belongs to a shuffle and a glide planes at once. On the basis of geometrical arguments, it has been suggested that this mixed shuffle-glide configuration is a saddle point, and therefore, is not stable \cite{Cel61JPCS}. This has been confirmed for Si by first principles calculations \cite{Piz03PMA}. Therefore, in this work, the B configuration is obtained by constraining the two atoms closest to the dislocation line to remain at the same distance along $[\bar{1}01]$. Finally, C is at the center of a short hexagon bond, at the intersection of two glide planes. The table~\ref{tabenergy} shows different energetic parameters, deduced from the first principles calculations, such as the energy differences between configurations, the c
 ore r
adii according to~\cite{Hir82WIL}, and the core energies, obtained for a core radius equal to the Burgers vector. Data for silicon, already presented in~\cite{Piz03PMA}, are included for comparison with other materials, and will not be discussed here at length. The most important point is the stability of the shuffle configuration A, both energy of B and C cores being substantially higher. 

\begin{table}
\caption{Energetic data calculated for the A, B and C configurations: total energy in eV/Burgers vector (relative to the most stable configuration, in box), core radius, and core energy (calculated with a core radius equal to the Burgers vector). Note that the configuration C is not stable for Ge and GaAs and relaxes to A.}\label{tabenergy}
\begin{center}

\begin{tabular}{@{\extracolsep{10pt}}*{10}{c}}
\hline
\hline
  & \multicolumn{3}{c}{Energy (eV/Bv)} & \multicolumn{3}{c}{Core radius (\AA)} & 
  \multicolumn{3}{c}{Core energy (eV.\AA$^{-1}$)} \\
  & A & B  & C & A & B & C & A & B & C \\
\hline
C & 1.29 & 2.51 & \fbox{0} & 0.61  & 0.46  & 0.84 & 2.33  &  2.82  & 1.82  \\
SiC & \fbox{0} & 0.43  & 0.29 & 0.74  & 0.62  & 0.66 & 1.21  & 1.35  &  1.30 \\
Si & \fbox{0} & 0.36 & 0.86 & 1.22 & 1.03  & 0.74 & 0.52 &  0.60 & 0.74 \\
Ge & \fbox{0} & 0.24  & $\rightarrow \textrm{A}$ & 1.11  &  0.95  &   & 0.49  & 0.55  &   \\
GaAs & \fbox{0} & 0.21  & $\rightarrow \textrm{A}$ & 1.11 &  0.95 &   & 0.44  &  0.49  &   \\
\hline
\hline
\end{tabular}
\end{center}\end{table}

In the case of diamond, it appears that the C core is clearly the most stable configuration, and B the less favored. The large energy difference with the A core may be explained by the narrowness of the C geometry, as well as the presence of two atoms in the core with a sp$^2$ hybridization, energetically favored for diamond. The bond length between these two atoms is 1.31~\AA, less than the first neighbor distance in graphite (1.42~\AA). Our results indicate that a non dissociated screw should be located in glide planes. However, it has been shown that dissociation is favored in this set of planes \cite{Blu03PRB}, and non dissociated dislocations are then unlikely to happen in diamond.  The determined energy ordering is in agreement with a recent study \cite{Blu03PRB}, but our computed energy differences are much lower, with $\Delta E_{\mathrm{AC}}=1.29$~eV/Bv and $\Delta E_{\mathrm{BC}}=2.51$~eV/Bv to compare with 3.3~eV/Bv and 5.4~eV/Bv respectively. The lower precision of
  the 
self-consistent tight-binding method used in~\cite{Blu03PRB} may explain this difference. 

Silicon carbide is interesting because it can be considered as an intermediate case between silicon and diamond. Deformation at high stress/low temperature of hexagonal 4H-SiC suggested the coexistence of undissociated and partial dislocations \cite{Dem05PSS}. However, apart from our previous semi-empirical approach, undissociated dislocations have not been investigated. Our calculations reveal that the A configuration is most stable like in silicon. The C configuration is second, with an energy difference much lower than for diamond. These results are in contrast with previous Tersoff potential calculations, giving C as the most stable core structure \cite{Piz02JPCM}. Screw configurations for silicon carbide include the same number of core C and Si atoms, due to the zinc-blende structure. 
Therefore, in the case of a C configuration, sp$^2$ hybridized core atoms are one carbon and one silicon, the former being energetically favored unlike the latter. 
We have tried to build a C configuration with two sp$^2$ carbon atoms, by exchanging the center Si atom with one of its carbon first neighbor. However, the resulting  structure is not favored, with an excess energy of 2.58~eV/Bv compared to the original C configuration. Non-stoichiometric core configurations have not been investigated, such a task being beyond the scope of the present study. 

Germanium and gallium arsenide show very similar behaviors regarding screw characteristics. In fact, only the A configuration is stable, and the energy differences $\Delta E_{\mathrm{AB}}$ are very close for both materials. Also, the A core was obtained after relaxation of an initial C configuration in both cases. Non dissociated screw dislocations have been observed in GaAs deformed at low temperature~\cite{Suz99PMA}. Our calculations unambiguously indicate that such dislocations are lying  in shuffle set planes. 

In the table~\ref{tabenergy}, we have reported core radii for each materials and configurations, determined from anisotropic elasticity theory according to the definition given in~\cite{Hir82WIL}. It is usually assumed that an adequate core radius value for covalent materials is one quarter of the Burgers vector. Here, we got a good agreement in the sole case of silicon carbide. Otherwise, the calculated core radius is usually 10-15\% (Ge and GaAs) to 30\% (Si and diamond) larger, showing the limitations of such an hypothesis.

\begin{figure}
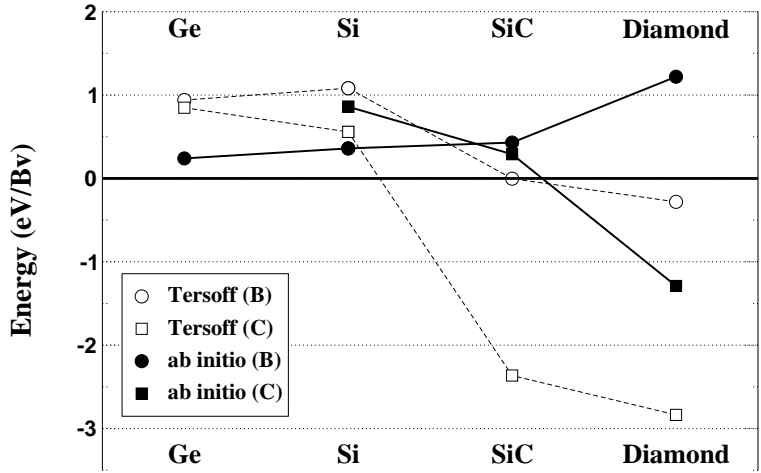

\onefigure[width=10cm]{fig2.eps}
\caption{Variation of screw configuration energies, for Ge, Si, SiC and diamond, calculated with the Tersoff potential and with a first principles method. The A core energy is the reference for each material.}
\label{energyIV}
\end{figure}

Elements in the same column of the periodic table are expected to share a number of properties. Here, it is then interesting to compare energy differences, calculated for group IV materials. Results from our first principles simulations are reported in the Figure~\ref{energyIV}, as well as previous Tersoff calculations~\cite{Piz02JPCM}.   
Going from Ge to diamond, the C core stability clearly increases, until C becomes the most stable configuration (diamond). A possible explanation is the presence of hybridized sp$^2$ atoms in the core. Such an electronic structure is highly unfavorable for Ge and 4s and 4p orbitals compared to sp$^3$ (C is unstable in that case), while it greatly lowers the core energy for diamond and 2s and 2p orbitals. Comparison between first principles and semi-empirical results indicates that the variation obtained with the Tersoff potential is qualitatively correct. The sp$^3\rightarrow\;$sp$^2$ change in the electronic structure is then adequately modeled by this potential. However, the energy variation is too steep, C instead of A being obtained as the most stable configuration for SiC. $\Delta E_{\mathrm{AB}}$, calculated from first principles, increases going from Ge to diamond, and the B configuration is always unstable. The B core is characterized by broken covalent bonds along the dislocation line, and the energy variation simply suggests that it is more expansive to break bonds for diamond than for Ge. Using the Tersoff potential, B is metastable for all materials, and the energy variation is reversed. This discrepancy confirms that semi-empirical potentials are generally unable to describe with a good accuracy covalent systems including dangling bonds, such as surfaces or vacancy-cluster defects. 

In conclusion, by performing first principles calculations, we have shown that characteristics of a non-dissociated screw dislocation are very close for most of the investigated materials, i.e. silicon, cubic silicon carbide, germanium, and gallium arsenide. The most stable configuration is obtained for a dislocation lying in shuffle set planes. Only diamond exhibits different properties, the lowest energy structure being obtained for a screw dislocation located in the glide set. The stability of this glide core structure, including sp$^2$ atoms, improves significantly from Ge to diamond, a characteristic fairly well reproduced with the Tersoff potential.

\acknowledgments

One of us, L.P., wants to thank A.~T.~Blumenau for fruitful discussions at the bar during the EDS2004 conference.


\end{document}